\documentclass[conference]{IEEEtran}

\usepackage[nolist]{acronym}
\usepackage[letterpaper, left=1in, right=1in, bottom=1in, top=0.75in]{geometry}
\begin{acronym} 
\acro{3GPP}{3rd Generation Partnership Project}
\acro{5G}{Fifth Generation}
\acro{4G}{Forth Generation}
\acro{3G}{Third Generation}
\acro{AoD}{Angle of Departure}
\acro{DoA}{Angle of Arrival}
\acro{BB}{Base Band}
\acro{BBU}{Base Band Unit}
\acro{BDM}{Beam Division Multiplexing}
\acro{BDMA}{Beam Division Multiple Access}
\acro{BER}{Bit Error Rate}
\acro{BS}{Base Station}
\acro{BW}{bandwidth}
\acro{C-RAN}{Cloud Radio Access Networks}
\acro{CSI}{Channel State Information}
\acro{CC}{Command and control}
\acro{DAC}{Digital-to-Analog Converter}
\acro{DAS}{Distributed Antenna Systems}
\acro{DBA}{Dynamic Bandwidth Allocation}
\acro{DL}{Downlink}
\acro{EE}{Energy Efficiency}
\acro{EIRP}{Equivalent Isotropic Radiated Power}
\acro{eNB}{Evolved Node B}
\acro{FEC}{Forward Error Correction}
\acro{FSO}{Free Space Optics}
\acro{GS}{Ground Station}
\acro{GSM}{Global System for Mobile Communications}
\acro{HAP}{High Altitude Platform}
\acro{HL}{Higher Layer}
\acro{HW}{Hardware}
\acro{HARQ}{Hybrid-Automatic Repeat Request}
\acro{IoT}{Internet of Things}
\acro{LAN}{Local Area Network}
\acro{LAP}{Low Altitude Platform}
\acro{LL}{Lower Layer}
\acro{LOS}{Line of Sight}
\acro{LTE}{Long Term Evolution}
\acro{LTE-A}{Long Term Evolution Advanced}
\acro{MAC}{Medium Access Control}
\acro{MAP}{Medium Altitude Platform}
\acro{ML}{Medium Layer}
\acro{Ml}{Machine Learning}
\acro{MME}{Mobility Management Entity}
\acro{MMSE}{Minimum Mean Square Error}
\acro{MRC}{Maximum Ratio Combiner}
\acro{mmWave}{millimeter Wave}
\acro{MIMO}{Multiple Input Multiple Output}
\acro{NFP}{Network Flying Platform}
\acro{NFPs}{Network Flying Platforms}
\acro{NGO}{Non-governmental Organization}
\acro{NLOS}{Non Line of Sight}
\acro{NN}{Neural Networks}
\acro{OLOS}{Obstructed Line of Sight}
\acro{OFDM}{Orthogonal Frequency Division Multiplexing}
\acro{PAM}{Pulse Amplitude Modulation}
\acro{PAPR}{Peak-to-Average Power Ratio}
\acro{PHY}{physical layer}
\acro{QAM}{Quadrature Amplitude Modulation}
\acro{QoE}{Quality of Experience}
\acro{QoS}{Quality of Service}
\acro{QPSK}{Quadrature Phase Shift Keying}
\acro{RF}{Radio Frequency}
\acro{RN}{Remote Node}
\acro{RRH}{Remote Radio Head}
\acro{RRC}{Radio Resource Control}
\acro{RRU}{Remote Radio Unit}
\acro{SCBS}{Small Cell Base Station}
\acro{SDN}{Software Defined Network}
\acro{SNR}{Signal-to-Noise Ratio}
\acro{SINR}{Signal-to-Interference-plus-Noise Ratio}
\acro{SON}{Self-organising Network}
\acro{TDD}{Time Division Duplex}
\acro{TD-LTE}{Time Division LTE}
\acro{TDM}{Time Division Multiplexing}
\acro{TDMA}{Time Division Multiple Access}
\acro{UE}{User Equipment}
\acro{UAV}{Unmanned Aerial Vehicle}
\acro{UDN}{Ultra Dense Network}
\acro{UL}{Uplink}

\end{acronym}  
\usepackage{cite}
\usepackage{xcolor}
\ifCLASSINFOpdf
\usepackage[pdftex]{graphicx}
\title{Saving Lives at Sea with UAV-assisted\\ Wireless Networks}
\author{\IEEEauthorblockN{Gianluca Fontanesi\IEEEauthorrefmark{1}, Anding Zhu\IEEEauthorrefmark{1}, and Hamed Ahmadi\IEEEauthorrefmark{1}\IEEEauthorrefmark{2}}
\\
\IEEEauthorrefmark{1}School of Electrical and Electronic Engineering, University College Dublin, Ireland\\
\IEEEauthorrefmark{2}School of Computer Science and Electronic Engineering, University of Essex, UK
\thanks{This publication has emanated from research conducted with the financial support of Irish Research Council under Grant GOIPG/2017/1741.}
}

\begin{document}
\maketitle

\begin{abstract}
In this paper, we investigate traits and trade-offs of a system combining \ac{UAV}s with \ac{BS} or \ac{C-RAN} for extending the terrestrial wireless coverage over the sea in emergency situations. Results for an over the sea deployment link budget show the trade-off between power consumption and throughput to meet the Search and Rescue targets.
\end{abstract}
\section{Introduction}
One  of  the  greatest  crisis in Europe today is the rising number of casualties among the immigrates through the Mediterranean sea. Since 2014,  more  than  14,500  people have lost their  lives  in  their  attempt  to  reach  Europe's  shores \cite{IOM1}. Emergency rescue operations are deadly needed in the Mediterranean area. However, 
due to lack of proper communication infrastructures, the rescue and recovery operations in such environment are limited.
The existing communications are mainly relying on the satellite service which has disadvantages such as high cost and long latency. Moreover, not all the smaller vessels and passengers own a satellite phone.
%
For these reasons, new technologies are being investigated to provide a reliable communication over the sea for enhancing timely rescue and recovery operations. One of the interesting solutions is to extend the coverage of the conventional terrestrial wireless networks to the sea to provide wireless connections to the people in needs via UAV-assisted networks.

\section{Related Work}
\ac{UAV}s will be used in next generation wireless networks. In the simplest form, \acp{UAV} are used as flying \ac{UE} or receive and forward relay for enhancing the connectivity of ground wireless devices\cite{Zhang_UAV_opportunitiesChallenges}. 
%

%
%

In addition to that \acp{UAV} are able to deploy or carry aerial \ac{BS}s to deliver wireless connectivity to desired areas. In particular, they can complement existing cellular systems in need of additional capacity, e.g. \ac{LTE}, 
or expand the cellular coverage and deliver internet services to remote and dedicated regions where infrastructure is not available or expensive to deploy \cite{Ahmadi_SON_airborne}.
Another attractive solution is \ac{UAV}s acting as \ac{RRH}\cite{Saad_CachingRRH_UAV}. 

In emergency networks applications \acp{UAV}, due to their properties, are a promising solution to satisfy the robustness, efficiency and rapidity requirements of Search and Rescue tasks. 
In \cite{Murphy_Sreenan_RescueMountainUAV} low cost \acp{UAV} are used in mountain terrain to survey and locate individuals who may be in distress. 
To the best of our knowledge, not enough attention has been given to \ac{UAV} as communication provider for Search and Rescue operations in the sea.

\section{System Description}

\begin{figure}
    \centering
    \includegraphics[width=1\columnwidth]{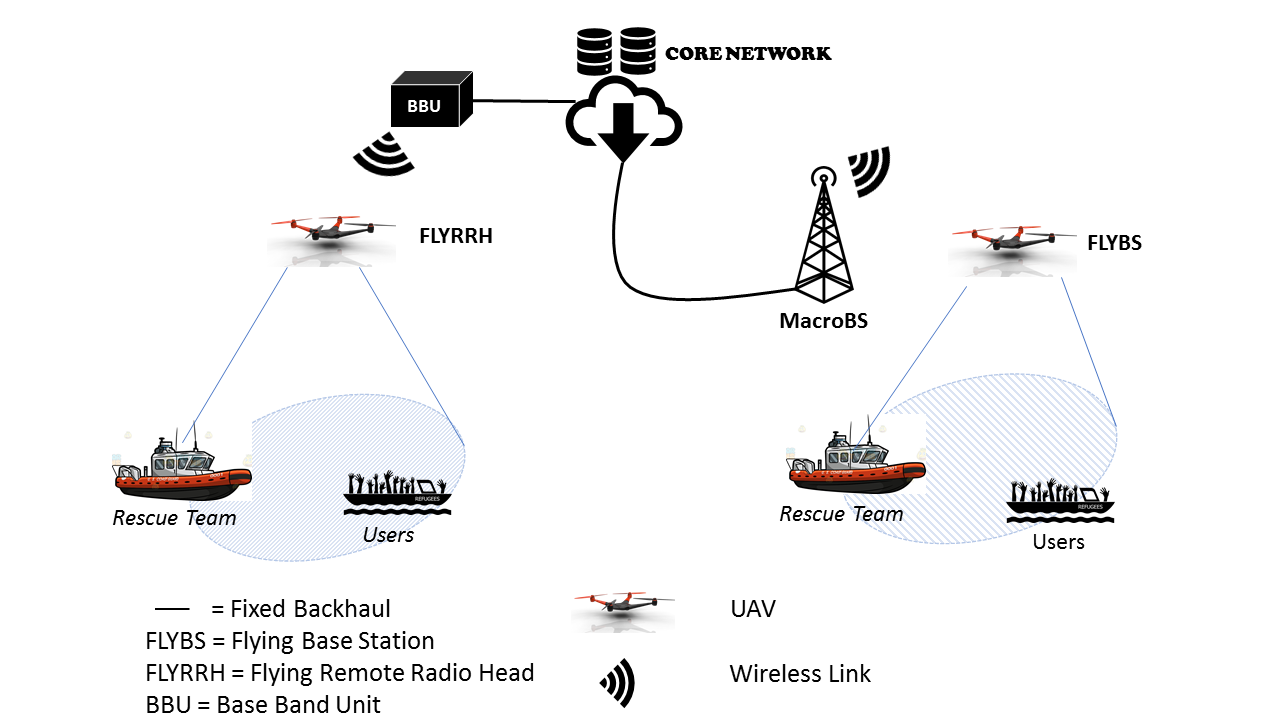}
    \caption{Graphical illustration of the architecture for extending the cellular coverage over the sea using flying BS and RRH}
    \label{fig:SystemMOdel}
\end{figure}

In case of an emergency, an airborne network can be used to provide coverage for mobile users and rescue operators in areas not covered by terrestrial network\cite{Ahmadi_SON_airborne}. 
In a sea scenario \ac{UAV}s could be deployed to act as flying \ac{BS} or flying \ac{RRH} to serve the users. In this way, \acp{UAV} would ensure access to the core network through a wireless backhaul from the flying \ac{BS}, or via wireless fronthaul link between the ground \ac{BBU} and the \ac{UAV}-\ac{RRH} (see Figure \ref{fig:SystemMOdel}). Moreover, the altitude of \acp{UAV} allows a direct connection to the users. As a first consequence, we consider for the \ac{UAV}-users and \ac{UAV}-cloud links \ac{LOS} connections. As second, we can operate at \ac{mmWave} band.
At these frequencies, taking into consideration an extra attenuation due to dry air and water vapor, the free space path loss model is the most accurate representation of the path loss occurring over the sea \cite{Khawaja_mmWave_A2G}.

\section{UAV as RRH or BS}
The deployment of UAV-assisted cellular networks struggle with issues at different levels. In particular, mechanical and power limitations affect the choice on the communication technology to be mounted on the aerial platform \cite{AerialNetowrk_ImplementationDesign}. 
Thus, a trade-off between throughput, complexity, weight and power consumption of the access solution must be investigated.

Flying cellular \ac{BS} can potentially provide high data rate wireless services to several users.
High throughput communication would enable applications like video streaming, but it would contribute also to an increase in the consumed power for data processing at the \ac{BS}. 
\ac{UAV}s have a limited amount of on-board energy which must be used for all its tasks. An increase in the volume of traffic reduces the energy available for hover and flight time, key resources for Search and Rescue operations over the sea.
The other side of the coin is that a limited transmission power restricts the mobility of the \ac{UAV}, that may need to assist rescue team and users in distress and reach the core network also at long distances from the land.
The deployment of antenna arrays on the \ac{UAV} can provide a useful gain to reduce the required transmission power at the front end, but the effect on the total consumed power must be investigated.
Hence, the trade off between power consumption and system throughput requires careful consideration as it can significantly impact the performance of the \ac{UAV} communication over the sea.


\ac{C-RAN} has been proposed to enhance resource management and reduce power consumption at the edge by centralising the baseband processing.
The implementation of \ac{RRH} on \ac{UAV} within a \ac{C-RAN} system can represent a promising approach that combines the benefits of an aerial communication with the ones given by a cloud architecture\cite{Saad_CachingRRH_UAV}.
Due to their compact size and lighter weight, \ac{RRH} platforms are more suitable for energy limited \acp{UAV}. In addition, all digital signal processing will be performed at BBU, requiring the \acp{UAV} lower computation power.
Moreover,
the power efficient design of \ac{RRH} would not require an active cooling system on the \ac{UAV}, leading to lower energy consumption.
All these advantages allow \acp{UAV} to save power in exchange of a longer activity without affecting the quality of the transmission.
Though this approach seems to encourage the deployment of \ac{UAV}-assisted networks for Search and Rescue operating over the sea, challenges remain.
A main challenge is the limited capacity of the wireless fronthaul, that leads to a trade off between number of \acp{UAV} for hovering, data rate and energy consumption.


\begin{table}
 \centering
 \caption{\footnotesize{Challenges Comparison}}\label{tab:Params}
    \footnotesize 
    \begin{tabular}{|p{0.2\columnwidth}|p{0.3\columnwidth}|p{0.2\columnwidth}|}
        \hline
        \emph{Challenge} & \emph{FLY-BS} & \emph{FLY-RRH}\\
        \hline  
        Power dissipation & Transmission Power, Computation Power & Transmission  Power\\
        \hline 
        Computation Power & at BBU & cloud/caching computation functions \\
        \hline 
        Load & more than 2kg  &  1 kg \\  
        \hline
        Limitation & Backhaul Capacity & Fronthaul Capacity \\
        \hline
        Processing & UAV & Cloud \\
        \hline
        Cooling & Passive & Passive \\
        \hline
    \end{tabular}
\end{table}

\section{Results}

\begin{figure}[ht]
    \centering
    \includegraphics[width=1\columnwidth]{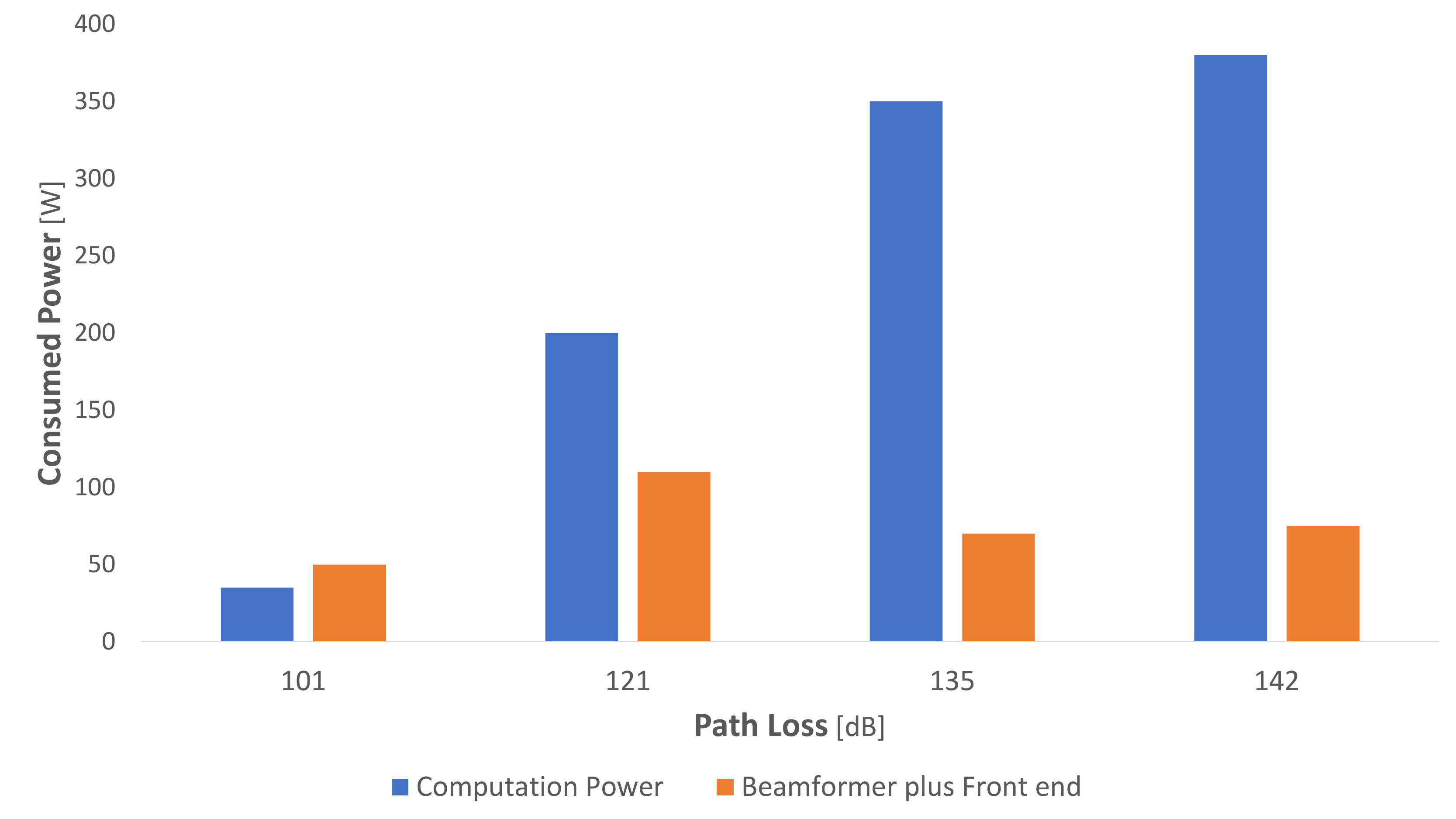}
    \caption{Power consumption of a \ac{UAV}-assisted wireless network at different distances from the core network}
    \label{fig:Results}
\end{figure}

We have simulated the target \ac{EIRP} required to overcome the path loss over the sea at \ac{mmWave} frequencies delivering an uplink throughput of 1 Gbps. For increasing path loss values, we have computed the array gain to be introduced on the \ac{UAV} to mantain the \ac{EIRP} and the resulting consumed power \footnote{values of computation power taken from \cite{SmallCell_EE}} (see Fig \ref{fig:Results}).
This work shows for the proposed approaches the trade off between the targets of mobility and data rate given a limited average output power and power consumption.

\section{Conclusion}

We investigated flying \ac{BS} and flying \ac{RRH} as promising architectures to provide wireless coverage over the sea for Search and Rescue operations. 
For the proposed approaches, we have evaluated the feasibility of satisfying needs and power constraints.


\bibliographystyle{IEEEtran}
{\footnotesize   
\bibliography{biblio.bib} 
}
\end{document}